 \documentclass[preprint2]{aastex}

\usepackage{mathrsfs}

\def\dsm{$M_\odot$}

\def\dra{$r_{01}$}
\def\drb{$r_{10}$}
\def\ddn{$\Delta\nu$}
\def\dhz{$\mu$Hz}
\def\dnmax{$\nu_{max}$}
\def\dnac{$\nu_{ac}$}
\def\dnu{$\nu_{n,l}$}
\def\dteff{$T_{\mathrm{eff}}$}

\shorttitle{Estimating The Mean Linewidth And Lifetime of Solar-like Oscillations}
\shortauthors{Wuming Yang}

\begin{document}


\title{Estimating The Mean Linewidth And Lifetime of Solar-like Oscillations of Stars}
\author{Wuming Yang}
\affil{Department of Astronomy, Beijing Normal University,Beijing 100875, China}
\email{yangwuming@bnu.edu.cn}

\begin{abstract}
\textbf{Scaling formulas were} deduced to describe the relations between 
the fundamental stellar parameters and the mean \textbf{linewidth and lifetime}
of solar-like oscillations of stars. The mean \textbf{linewidth and} lifetime 
of solar-like oscillations \textbf{are dependent on the large frequency separation, 
the effective temperature, and the acoustic impedance ($\rho c$) in the photosphere} 
of stars. The mean lifetime \textbf{can be} approximate to the lifetime of the mode 
with $\nu\sim\nu_{max}$. We compared the results of the scaling relations with the mean 
lifetimes of solar-like oscillations of stars observed by \textit{Kepler} 
and \textit{CoRoT}, which shows that the observed mean lifetimes are reproduced 
well by the scaling relations. The dependence of the mean lifetime on \textbf{the} large 
frequency separation, the effective temperature, and the acoustic impedance of stars 
\textbf{indicates} that lifetimes of solar-like oscillations rely on the mass and evolutionary 
phase of stars. Moreover, our calculations show that the mean lifetimes of $p$-modes 
of stars can be affected by metallicity abundances. 
\end{abstract}


\section{INTRODUCTION}
By matching the oscillation frequencies of models with the observed ones,
asteroseismology is used to determine the fundamental parameters of stars, 
such as mass, radius, and age, and diagnose internal structures 
of stars \citep{roxb03, chri10, dehe10, chap14, 
guen14, liu14, yang15, gugg16, rodr17, silv17}. Asteroseismology
plays an irreplaceable role in studying the structure and evolution of stars. 
The measurable characteristics of solar-like oscillations of stars mainly
include individual frequencies \dnu{}, the large frequency separation \ddn{},
the frequency of maximum power \dnmax{}, and so on. 

Individual frequencies are widely used in asteroseismology.
In order to determine the fundamental parameters of stars, the observed 
frequencies \dnu{} are directly compared with those calculated from models 
by the chi-squared method or the maximum likelihood function method.
By calculating the ratios of the small separations to the large 
separations, \dra{} and \drb{}, from individual frequencies,
individual frequencies are also used to extract the information 
about the internal structures of stars \citep{roxb03, roxb07, cunh07,
dehe10, liu14, yang16b}. 

The large frequency separation, \ddn{}, and the frequency of maximum power, \dnmax{},
are \textbf{considered to be} more easily-measured than individual frequencies. 
The value of \ddn{} is proportional to the mean density of stars \citep{ulri86}. 
The value of \dnmax{} scales with the acoustic cutoff frequency (\dnac{}) \citep{brow91}. 
Basing on these results, \cite{kjel95} obtained the well-known scaling relations:
\begin{equation}
\Delta\nu=(M/M_{\odot})^{1/2}(R/R_{\odot})^{-3/2}\Delta\nu_{\odot},
\label{scal1}
\end{equation}
and
\begin{equation}
\nu_{max}=\frac{M/M_{\odot}}{(R/R_{\odot})^{2}\sqrt{T_{\rm eff}/5777}}\nu_{max,\odot}.
\label{scal2}
\end{equation}
The scaling relations are extensively used to estimate masses and radii of stars and 
study the characteristics of star populations \citep{migl09, yang10, shar17}. 

Besides the frequencies, the amplitude and the FWHM linewidth $\Gamma$
of $p$-modes of stars also can be determined \citep{chap09, hekk10, baud11, appo14}. 
The lifetime $\tau$ is considered to be related to the linewidth $\Gamma$ 
by $\Gamma=1/(\pi\tau)$. \cite{chap09} is the first to try to find a simple
scaling relation to describe the average lifetime, $\langle\tau\rangle$, 
of solar-like oscillations and obtain $\langle\tau\rangle\propto T_{\rm eff}^{-4}$. 
Hereafter, we throw out the average symbol $\langle \rangle$.
\cite{baud11} obtain $\tau\propto T_{\rm eff}^{-s}$, where the value of 
the $s$, however, is $14\pm8$ for main-sequence (MS) stars and $-0.3\pm0.9$ 
for red giants. The difference between the result of Chaplin and that of
Baudin is very significant. The lifetime is potentially an extremely 
useful diagnostic of near-surface convection in stars, affects 
the detectability of modes, and aids in better understanding the excitation
and extinction mechanisms of modes, so that the lifetime
predictions play an important role in asteroseismology \citep{chap09}.

In this work, we give \textbf{some simple scaling relations to describe
the average linewidth $\Gamma$ and} lifetime $\tau$ of solar-like oscillations. 
The paper is organized as follows: in Section 2, we deduce the scaling relations
and compare the results of the scaling relations with observations, then 
discussion and summary are given in Section 3.


\section{SCALING RELATIONS OF MEAN LINEWIDTH AND LIFETIME OF SOLAR-LIKE OSCILLATIONS}
\subsection{Scaling Relation of Mean Linewidth of Oscillations}

The energy transported per unit time across a spherical surface by
propagating acoustic waves with angular frequency $\omega$ is 
\begin{equation}
F=4\pi r^{2}\frac{1}{2} \rho |\mathbf{v}|^{2}c,
\end{equation}
where $\mathbf{v}$ is the velocity of oscillations, $c$ the adiabatic sound
speed, $\rho$ the density, $r$ the radius. The total energy of
the oscillations in a star is \citep{balm90}
\begin{equation}
\begin{array}{ll}
E_{osc}&=2\int^{R}_{0}\frac{1}{2}\rho |\mathbf{v}|^{2} 4\pi r^{2}dr \\
       &= 2F\int^{R}_{0}\frac{dr}{c}\\
       &= F\Delta\nu^{-1}.
\end{array} 
\label{p1}
\end{equation}

The growth rate $\omega_{i}$ of amplitude of an oscillation, which 
is the imaginary part of $\omega(\omega=\omega_{r}+i\omega_{i})$, 
is related to the work integral $W$, i.e. 
\begin{equation}
\omega_{i}=-\frac{1}{2} \frac{W/E_{osc}}{\Pi},
\label{damp1}
\end{equation}
where the period $\Pi=2\pi\omega_{r}^{-1}$, and the work integral $W$
is defined as (see equations 25.16, 25.17 and 25.18 of \cite{unno89}
for more details of $W$ and $E_{osc}$)
\begin{equation}
W =\oint dt \frac{dE_{osc}}{dt}.
\label{w1}
\end{equation}
Therefore, we obtain
\begin{equation}
\omega_{i}=-\frac{\omega_{r}}{4\pi} \frac{W\Delta\nu}{F}.
\label{damp2}
\end{equation}

The work integral $W$ can be decomposed into $W_{N}$ and $W_{E}$,
which are related to the perturbations of nuclear energy generation rate and
energy transfer, respectively. Let us neglect $W_{N}$. We only consider radial 
oscillations in this work. The work integral $W_{E}$ is given as \textbf{
(see Equations 26.3 and 26.4 of \cite{unno89}, and Equation 3 of \cite{gold91})}
\begin{equation}
W_{E} =\frac{\pi}{\omega_{r}}\int^{R}_{0}dr(-\frac{\partial\Delta L}{\partial r})\frac{\Delta T}{T},
\label{we1}
\end{equation}
\textbf{where $L$ is luminosity and T temperature of a star. Let us estimate the order of magnitude 
of the integral of Equation (\ref{we1}) from adiabatic oscillations. 
For the adiabatic radial oscillations, 
\cite{gold91} gave 
\begin{equation}
\frac{\Delta T}{T} = -(\Gamma_{3}-1)\frac{\partial\xi}{\partial r} \sim -(\Gamma_{3}-1)\frac{\omega^{2}\xi}{g},
\label{dt1}
\end{equation}
where $\xi$ is radial displacement eigenfunction, $g$ gravity, and 
\begin{equation}
\Gamma_{3}-1 \equiv (\frac{\partial\ln T}{\partial\ln\rho})_{s}.
\end{equation}
Neglecting the change in $(\Gamma_{3}-1)$, \cite{gold91} gave 
\begin{equation}
\frac{\Delta T}{T} \sim -\frac{\omega^{2}\xi}{g},
\label{dt2}
\end{equation}
and 
\begin{equation}
\frac{\partial\Delta L}{\partial r} \sim -\frac{\Delta L}{H} \sim -\frac{L\omega^{2}\xi}{gH},
\label{dl1}
\end{equation}
where $H$ is the local pressure scale height. Then \cite{gold91} obtained
\begin{equation}
 \int^{R}_{0}dr(-\frac{\partial\Delta L}{\partial r})\frac{\Delta T}{T} \sim -\frac{L\omega^{4}\xi^{2}}{g^{2}}.
\end{equation}
}

\textbf{
Therefore, we can obtain
\begin{equation}
W_{E}\sim -\frac{\pi}{\omega_{r}}\frac{L\omega^{4}\xi^{2}}{g^{2}}.
\label{we2} 
\end{equation}
Substituting $W$ in Equation (\ref{damp2}) by $W_{E}$,} we obtain that the thermal damping
rate can be given as
\begin{equation}
\begin{array}{ll}
\omega_{i}& \sim\frac{1}{4} \frac{\Delta\nu}{F} \frac{L\omega^{4}\xi^{2}}{g^{2}}\\
         &=\frac{\sigma\pi r^{2}\omega^{4}\xi^{2}}{F g^{2}}\Delta\nu T^{4},
\end{array} 
\label{damp3}
\end{equation}
where $\sigma$ is Stefen constant.
For turbulent stresses, the work integral is given as \citep{gold91}
\begin{equation}
\begin{array}{ll}
W_{T}  & = \frac{\pi}{\omega_{r}}4\pi\omega^{2}\int^{R}_{0}dr r^{2}\rho\nu_{H}(\frac{\partial\xi}{\partial r})^{2} \\
       & \sim \frac{\pi}{\omega_{r}}\frac{L\omega^{4}\xi^{2}}{g^{2}},
\end{array} 
\label{wt}
\end{equation}
where $\nu_{H}$ is turbulent viscosity. Therefore, mechanical damping rate also has the 
expression of Equation (\ref{damp3}).

Solar-like oscillations are considered to be stable. The value of $F$ can be estimated as
\begin{equation}
F \sim 4\pi r^{2}\frac{1}{2} \rho c \omega^{2} \xi^{2}.
\label{flux}
\end{equation}
\textbf{Substituting $F$ in Equation (\ref{damp3}) by Equation (\ref{flux}), we obtain 
\begin{equation}
\omega_{i} \approx \frac{\sigma}{2\rho c} \frac{\omega^{2}}{g^{2}} \Delta\nu T^{4}
\label{damp4}
\end{equation}
where $\rho$, $c$, $g$, and $T$ take their values at $r=R$.
}

\begin{figure}
\centering
\includegraphics[scale=0.2]{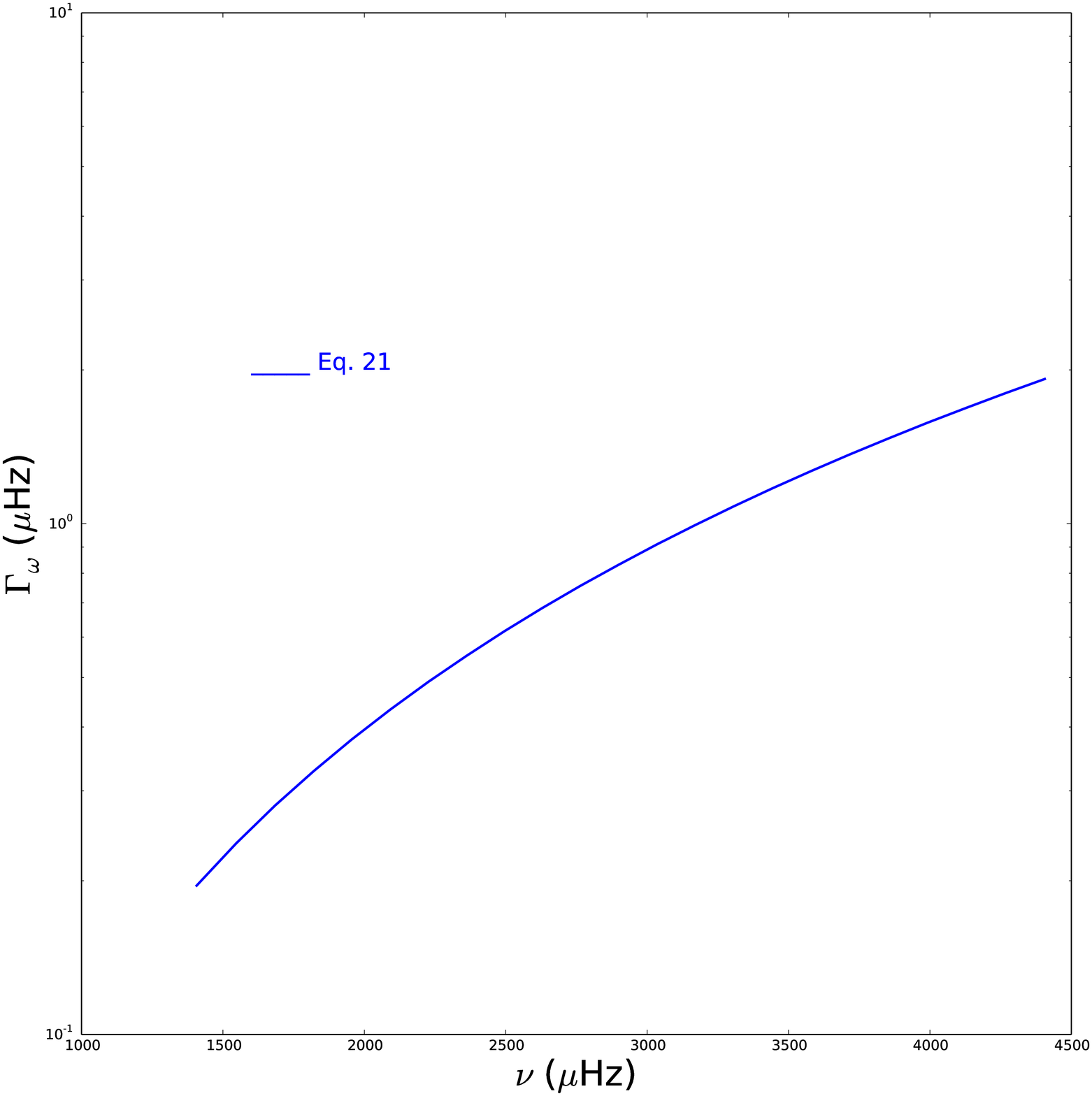}
\caption{
The linewidth of radial modes of the Sun as a function of frequency, 
calculated from GS98M \citep{yang16a} by using Equation (\ref{gama3}). 
The values of $\rho(R)$ and $c(R)$ of GS98M are $1.516\times10^{-7}$ g cm$^{-3}$
and $7.908\times10^{5}$ cm s$^{-1}$, respectively.}
\label{fig1}
\end{figure}

\textbf{ 
The linewidth is considered to be related with the damping rate by $\Gamma_{\omega}=\omega_{i}/2\pi$, i.e.
\begin{equation}
\Gamma_{\omega} \approx \frac{\sigma}{4\pi\rho c} \frac{\omega^{2}}{g^{2}} \Delta\nu T^{4}.
\label{gama1}
\end{equation}
Due to the fact that frequencies $\nu_{n,l}$ have an approximately equal \ddn{}, the mean linewidth
$\Gamma$ is approximate to $\Gamma_{\omega_{max}}$, i.e.
\begin{equation}
\Gamma \simeq \Gamma_{\omega_{max}} \approx \frac{\sigma}{4\pi\rho c} \frac{\omega^{2}_{max}}{g^{2}} \Delta\nu T^{4}.
\label{gama2}
\end{equation}
The mean linewidth of the solar $p$-modes is about $0.95\pm0.08$ \dhz{} \citep{baud11} or
around $1.15\pm0.07$ \dhz{} \citep{chap09}. The value of $\Gamma$ of Equation (\ref{gama2}) is about $2.83$
\dhz{} at $\nu=3090$ \dhz{} for the Sun. This indicates that Equations (\ref{gama1}) and (\ref{gama2})
overestimate the linewidth of the Sun by about $3$ times. Equation (\ref{gama1}) divided by $3$, 
we obtained that the linewidths of solar-like oscillations can be estimated as
\begin{equation}
\Gamma_{\omega} \simeq \frac{\sigma}{12\pi\rho c} \frac{\omega^{2}}{g^{2}} \Delta\nu T^{4}.
\label{gama3}
\end{equation}
Figure \ref{fig1} shows the $\Gamma_{\omega}$ of the Sun as a function of frequency calculated by using 
Equation (\ref{gama3}), which indicates that the order of magnitude of linewidths of $p$-modes with $\nu\sim$
\dnmax{} can be properly estimated by Equation (\ref{gama3}).
}

\textbf{
Equation (\ref{dt2}) neglects the effect of $(\Gamma_{3}-1)$. 
Calculations show that the value of $(\Gamma_{3}-1)$ is less than $1$ at $r=R$. 
The more massive the star, the closer to $1$ the value of $(\Gamma_{3}-1)_{R}$. 
The lower the metallicity, the larger the value of $(\Gamma_{3}-1)_{R}$. 
The value of $(\Gamma_{3}-1)_{R}$ of MS models with $M\lesssim1.1$ \dsm{}
is approximate to that of the Sun. Therefore, linewidths of oscillations of
these stars can be estimated by Equation (\ref{gama3}). But the value 
of $(\Gamma_{3}-1)_{R}$ of MS models with masses between $1.1$ and $1.5$ \dsm{} 
is about $1-5$ times as large as that of the Sun, which is dependent on the mass, 
chemical compositions, and evolutionary stage of stars. Thus linewidths
of oscillations of these stars can be estimated by Equations (\ref{gama1})
and (\ref{gama2}). For these stars, Equation (\ref{gama1}) can be rewritten as
\begin{equation}
\Gamma_{\omega} \simeq f\frac{\sigma}{12\pi\rho c} \frac{\omega^{2}}{g^{2}} \Delta\nu T^{4},
\label{gama4}
\end{equation}
where the value of $f$ is mainly in the range of $1-5$. For the models with masses 
between $1.2$ and $1.5$ \dsm{}, the value of $(\Gamma_{3}-1)_{R}$ is mainly 
around $3$ times greater than that of the Sun. As a consequence, the mean value 
of $f$ is around $3$. Equations (\ref{dt2}), (\ref{dl1}), and (\ref{flux}) only 
give an estimation of the order of magnitude. Thus the value of $f$ is not only 
affected by $(\Gamma_{3}-1)_{R}$. The value of $f$ can be determined from detailed 
asteroseismic analyses of some F-type stars with solar-like oscillations.
}

\textbf{
Scale from the solar value to estimate the mean linewidth and use
\begin{equation}
\omega_{max}=\frac{g/g_{\odot}}{\sqrt{T_{\rm eff}/5777}}\omega_{max,\odot},
\label{scal22}
\end{equation}
one can obtain
\begin{equation}
\left\{
 \begin{array}{lc}
 \Gamma \simeq\frac{(\rho c)_{\odot}}{\rho c}\frac{\Delta\nu}{\Delta\nu_{\odot}}(\frac{T_{\rm eff}}{5777})^{3}\Gamma_{\odot},
                                                                                  &\mathrm{for } M\lesssim1.1 M_{\odot};\\
 \Gamma \simeq f\frac{(\rho c)_{\odot}}{\rho c}\frac{\Delta\nu}{\Delta\nu_{\odot}}(\frac{T_{\rm eff}}{5777})^{3}\Gamma_{\odot},
                                                                                  &\mathrm{for } M\gtrsim1.2 M_{\odot},\\
\end{array}\right.
\label{scga1}
\end{equation}
where the value of $\Delta\nu_{\odot}$ is $134.6$ \dhz{}, and $(\rho c)_{\odot}\simeq 0.12$ g cm$^{-2}$ s$^{-1}$. 
The value of $\Gamma_{\odot}$ is about $1.15$ \dhz{} \citep{chap09} or $0.95\pm0.08$ \dhz{} \citep{baud11}. 
The value of $f$ is mainly in the range of $1-5$, which is dependent on the mass and evolutionary stage of stars. 
Its mean value is around $3$.
}

\subsection{Scaling Relation of Mean Lifetime of Oscillations}

The \textbf{lifetime of a mode is} considered to be relevant to \textbf{the linewidth} 
of the mode \textbf{by $\tau_{\omega}=1/(\pi\Gamma_{\omega})$}. The average lifetime, 
$\tau$, of low-degree $p$-modes of the Sun given by \cite{chap09} is $3.2\pm0.2$ days, 
which is close to that calculated from the $\Gamma_{\omega}$ of the mode with $\nu \approx$ \dnmax{}.
Therefore, we obtain that the mean lifetime of $p$-modes can be estimated \textbf{as
\begin{equation}
\tau \approx \tau_{\omega_{max}} \simeq \left\{
 \begin{array}{lc}
 \frac{12\rho c }{\sigma} \frac{g^{2}}{\omega^{2}_{max}} \Delta\nu^{-1} T^{-4},&\mathrm{for } M\lesssim1.1 M_{\odot};\\
 \frac{12\rho c }{f\sigma} \frac{g^{2}}{\omega^{2}_{max}} \Delta\nu^{-1} T^{-4},&\mathrm{for } M\gtrsim1.2 M_{\odot}.\\
\end{array}\right.
\label{tau0}
\end{equation}
}

In order to estimate the mean lifetime of $p$-modes of stars, we scale from the solar value 
to \textbf{calculate the mean lifetime:
\begin{equation}
\tau \approx\left\{
 \begin{array}{lc}
 \frac{\rho c }{(\rho c)_{\odot}}\frac{\Delta\nu_{\odot}}{\Delta\nu}(\frac{5777}{T_{\mathrm{eff}}})^{3}\tau_{\odot},
                                                                           &\mathrm{for } M\lesssim1.1 M_{\odot};\\
 \frac{\rho c }{f(\rho c)_{\odot}}\frac{\Delta\nu_{\odot}}{\Delta\nu}(\frac{5777}{T_{\mathrm{eff}}})^{3}\tau_{\odot},
                                                                           &\mathrm{for } M\gtrsim1.2 M_{\odot},\\
\end{array}\right.
\label{tau1}
\end{equation} 
where the value of $\tau_{\odot}$ is about $3.2$ days \citep{chap09}
or $3.8$ days \citep{baud11}.}
This relation \textbf{shows} that the mean lifetime of solar-like oscillations
increases with a decrease in \ddn{} and \dteff{}. 

For MS stars and red giants, the value of $\rho(R)$ generally decreases 
with an increase in $R$. The increase in $M$ can result in an increase in radius,
i.e. can lead to a decrease in $\rho(R)$. Therefore, the acoustic impedance $\rho c$
generally decreases with an increase in mass and radius of stars. But for the same type stars, 
they are expected to have an \textbf{approximate $\rho c$. The parameter $f$ in 
Equation (\ref{tau1}) can be roughly replaced by the mean value $3$.}
As a consequence, for stars extracted \ddn{} and \dteff{}, their $\tau$ can be 
estimated from Equation (\ref{tau1}).

\textbf{For theoretical calculations, the value of $\rho c$ can be obtained from stellar models.
But for observations, the value of $\rho c$ of stars is hard to be obtained.}
The density $\rho(R)$ and sound speed $c$ decrease with mass and age of stars, 
i.e. decrease with an increase in mass and radius. Assume $\rho c \propto 1/(MR)$, 
Equation (\ref{tau1}) can be rewritten as
\begin{equation}
 \tau\simeq f_{2} \frac{\Delta\nu_{\odot}}{\Delta\nu}(\frac{5777 \mathrm{K}}{T_{\mathrm{eff}}})^{3} \frac{M_{\odot}R_{\odot}}{MR}\tau_{\odot},
 \label{tau2}
\end{equation}
where $f_{2}$ is a free nondimensional parameter. \textbf{
It is more convenient to estimate $\tau$ of stars by using Equation (\ref{tau2}). }

\textbf{
If the mass of a star is unknown, Equation (\ref{tau2}) is also hard to be used in observations. }
The masses of stars with solar-like oscillations are mainly between about $1$ and 
$1.5$ \dsm{}. The change \textbf{of the $R$ in Equation (\ref{tau2}) can be much
larger than that of the $M$.} Therefore, for stars whose radius is determined, their $\tau$ can 
be \textbf{approximately estimated as}
\begin{equation}
 \tau\simeq f_{3} \frac{\Delta\nu_{\odot}}{\Delta\nu}(\frac{5777 \mathrm{K}}{T_{\mathrm{eff}}})^{3} \frac{R_{\odot}}{R}\tau_{\odot},
 \label{tau3}
\end{equation}
where $f_{3}$ is a free parameter. \textbf{Our calculations show that the value 
of $f_{3}$ is $1$ for MS stars with $M\lesssim1.1$ \dsm{} and red giants 
(\ddn{} $\lesssim 40$ \dhz{}) and $1/3$ for MS stars with $M\gtrsim1.2$ \dsm{}.}
For stars observed $R$, \ddn{}, and \dteff{}, it is convenient to \textbf{estimate}
$\tau$ or $\Gamma$ by using Equation (\ref{tau3}). \textbf{But Equation (\ref{tau3}) 
is an approximation of Equation (\ref{tau2}).}

\subsection{Comparison with Observations}

\cite{lund17} determined the values of FWHM ($\Gamma_{\alpha}$) at \dnmax{} of $66$ 
stars (LEGACY Sample) from the observations of \textit{Kepler}. 
Moreover, \cite{hekk10} studied in detail the lifetimes of $p$-modes of 
four red giants from the observations of \textit{CoRoT} and gave the values
of radii of the four red giants. The values of $\tau$ of LEGACY Sample and 
\cite{hekk10} sample are obtained by $\tau=1/(\pi\Gamma)$. 
In addition, \cite{chap09} and \cite{baud11} also gave $\tau$ or $\Gamma$ 
of some stars. Thanks to these works, we have a large enough sample to 
test Equations (\ref{tau1}), (\ref{tau2}), and (\ref{tau3}).

\begin{figure}
\centering
\includegraphics[scale=0.5, angle=-90]{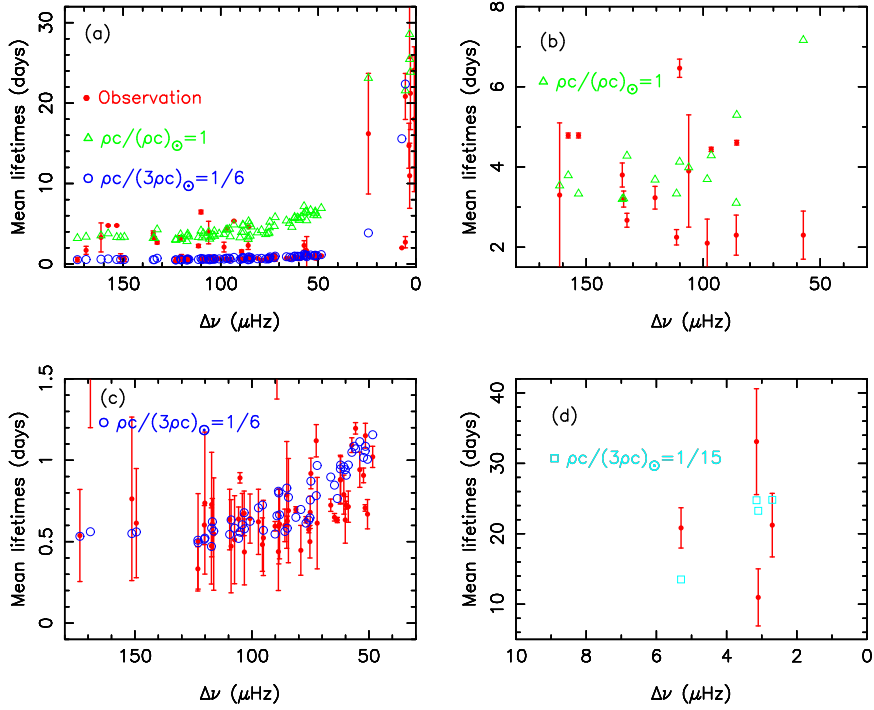}
\caption{
Average lifetimes of $p$-modes as a function of \ddn{}. The red dots represent the observations 
of \cite{lund17}, \cite{hekk10}, \cite{chap09}, and \cite{baud11}. The triangles, circles,
and squares refer to the results of Equation (\ref{tau1}) with $\tau_{\odot}=3.2$ days.
The value of $f$ is $3$ for `more massive' and `red' samples.} 
\label{fig2}
\end{figure}

The observed $\tau$ of the sample is shown in panel $a$ of Figure \ref{fig2}.
The results of Equation (\ref{tau1}) are also shown in the panel. Due to the 
fact that the values of $\rho c$ of the stars are unknown, we assumed that 
the value of $\rho c/(\rho c)_{\odot}$ is a constant in the calculations. 
The sample can be divided into three subsamples: one has $\tau> 2$ days and \ddn{} $>50$ \dhz{},
labeled as `low-mass' sample; one has $\tau < 2$ days, labeled as `more massive'
sample; four red giants of \cite{hekk10} are labeled as `red' sample.

Panel $b$ of Figure \ref{fig2} shows the mean lifetimes of the `low-mass' sample. 
Part of the sample is reproduced by Equation (\ref{tau1}) with $\rho c/(\rho c)_{\odot}=1$; 
but part cannot be reproduced correctly. This can be due to the assumption 
of $\rho c/(\rho c)_{\odot}=1$. For stars with mass less than $1$ \dsm{}, 
their $\rho c/(\rho c)_{\odot}$ may be larger than $1$; but for stars with 
mass larger than $1$ \dsm{}, their $\rho c/(\rho c)_{\odot}$ may be less than $1$.

Panel $c$ of Figure \ref{fig2} shows the mean lifetimes of the `more massive' sample. 
Most of the sample are reproduced well by Equation (\ref{tau1}) with $\mathbf{\rho c/[3(\rho c)_{\odot}]=1/6}$, 
which indicates that \textbf{the acoustic impedance ($\rho c$) of these stars is lower 
than that of the Sun. Because the more massive the stars, the smaller the $\rho c$,
masses of these stars} may be larger than $1$ \dsm{} and that the difference 
in the masses \textbf{may not be} significant, especially for the F-like stars. 

\begin{figure}
\centering
\includegraphics[scale=0.5, angle=-90]{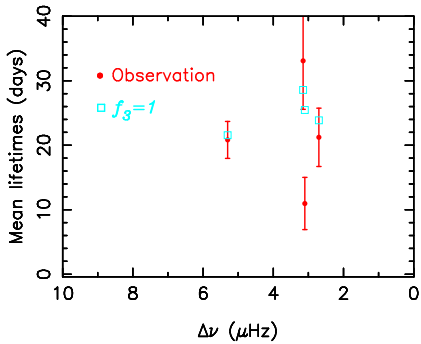}
\caption{
Mean lifetimes of red giants of \cite{hekk10} as a function of \ddn{}. The squares represent
the results of Equation (\ref{tau3}) with $\tau_{\odot}=3.2$ days and $f_{3}=1$. 
Three of the four are reproduced within a standard error.
}
\label{fig3}
\end{figure}

Panel $d$ of Figure \ref{fig2} shows the mean lifetimes of `red' sample of \cite{hekk10}. 
In order to reproduce the sample, the value of $\rho c/[3(\rho c)_{\odot}]$ 
decreases from $\mathbf{1/6}$ for `more massive' sample to $\mathbf{1/15}$. 
However, the results are unsatisfied.
This may be due to neglecting \textbf{the variation in $\rho c$ of red giants.}
The variation in radii of red giants is significant, \textbf{which leads to 
the fact that there is a significant difference in $\rho c$ of red giants.}
This hints to us that Equation (\ref{tau3}) could be more suitable for red giants.
Figure \ref{fig3} shows that the values of $\tau$ of the red giants are reproduced well 
by Equation (\ref{tau3}) with $f_{3}=1$. Three of the four are reproduced within 
a standard error. This indicates that scaling relations (\ref{tau1}) and (\ref{tau3}) 
work well. 

\subsection{Mean Lifetimes Calculated from Stellar Models}
 
Equation (\ref{tau1}) provides us with an opportunity to fastly calculate
the mean lifetimes of $p$-modes from stellar models. By using the Yale Rotation 
Evolution Code \citep[YREC]{pins89, yang07} in its nonrotation configuration, 
we calculated a serial stellar models with different masses and metallicities.
Mean lifetimes of oscillations of the models were calculated by using 
Equation (\ref{tau1}).

\begin{figure}
\includegraphics[scale=0.24]{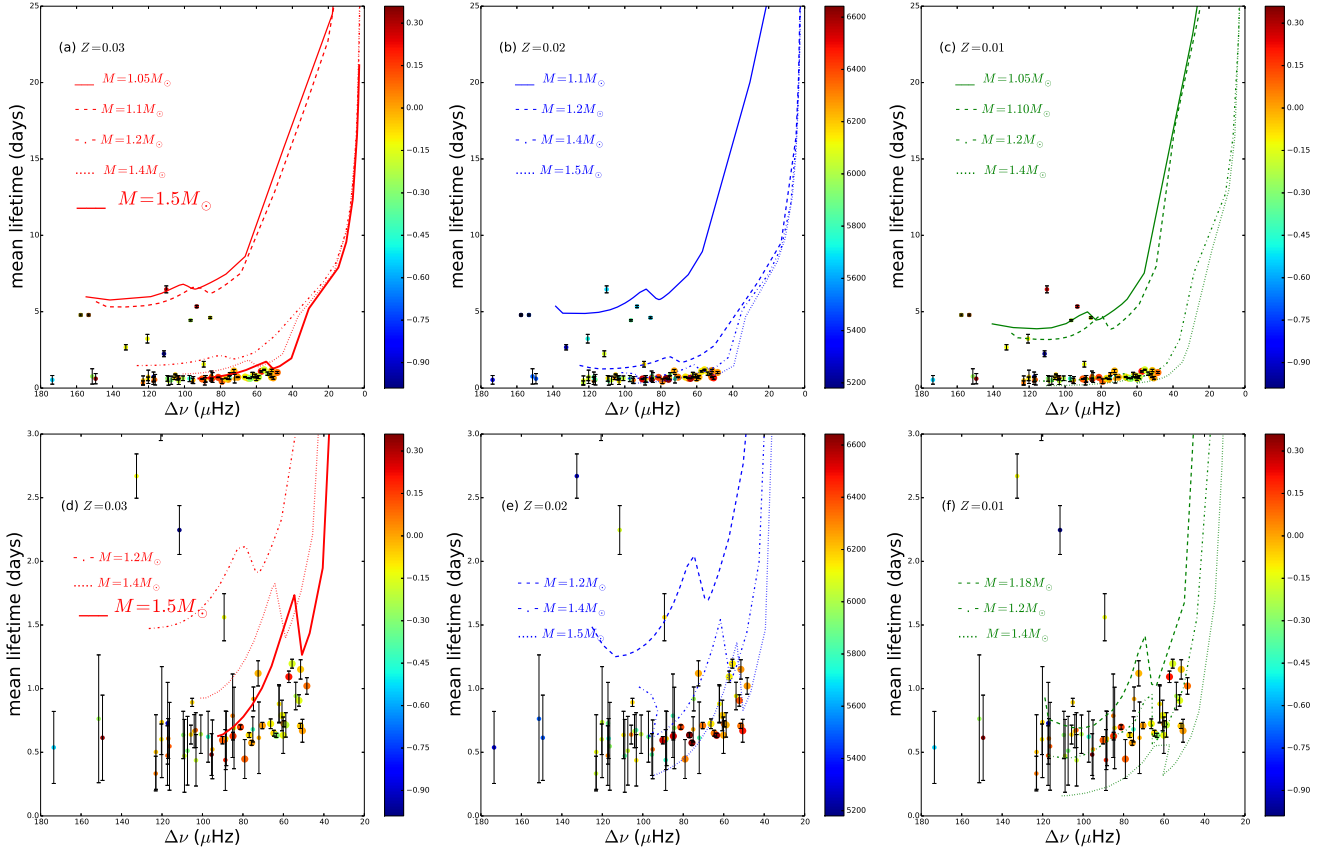}
\caption{ 
Comparison between the mean lifetimes of LEGACY sample \citep{lund17} and the
mean lifetimes calculated by using Equation (\ref{tau1}) with $\tau_{\odot}=3.8$ days
and $f=3$. 
The large circles represent F-like stars of LEGACY sample, while the small circles 
refer to Simple stars of LEGACY sample. The colorbars of panels $a$, $c$, $d$, and 
$f$ are proportional to [Fe/H] of LEGACY sample, while those of panels $b$ and $e$ 
are proportional \dteff{} of the sample.
}
\label{fig4}
\end{figure}

\begin{figure}
\centering
\includegraphics[scale=0.25]{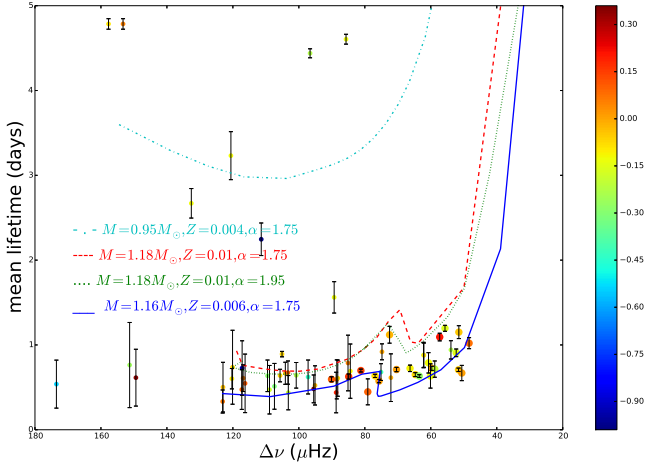}
\caption{ 
Same as Figure \ref{fig4}, but for the models with different masses and metallicities.
The symbol $\alpha$ is the mixing-length parameter.
}
\label{fig5}
\end{figure}

Calculated results are represented in \textbf{Figures \ref{fig4} and \ref{fig5}. 
The mean lifetimes of oscillations of F-like stars of LEGACY sample can be 
reproduced well by those of MS and MS-turnoff }
models with masses between about $\mathbf{1.2}$ and $1.5$ \dsm{}. 
\textbf{But} the masses of Simple stars of the sample change from 
about $\mathbf{1.0}$ to $\mathbf{1.2}$ \dsm{}. \textbf{The F-like stars are 
more massive, thus they have a higher effective temperature.}
The more massive the stars, the smaller the $\tau$. \textbf{The higher the metallicity, 
the larger the $\tau$ of MS stars. This is because the higher the metallicity,
the lower the luminosity of a star, and the lower the effective temperature. For
MS models with a given mass and radius, they have a given \ddn{}; the higher the metallicity,
the larger the value of $\rho c$, and the lower the effective temperature. As a consequence, 
the higher the metallicity, the larger the $\tau$. }

\textbf{Figures \ref{fig4} and \ref{fig5} show that post-MS models with $M\lesssim1.1$ 
have a $\tau$ larger than about $4$ days between \ddn{} $=80$ and \ddn{} $=60$ \dhz{}.
These stars could not have evolved into the post-MS stage. Thus we could hardly find 
the stars with $\tau>4$ days and \ddn{} between $80$ and $60$ \dhz{}. The lower 
the mass, the larger the $\tau$; the higher the metallicity, the larger the $\tau$.
This} explains why the Simple stars with $\tau \sim 5$ days have a higher metallicity
and a lower effective temperature (see panels $a$ and $b$ of Figure \ref{fig4}).
Moreover, Figures \ref{fig4} and \ref{fig5} show that the effect of metallicity on 
lifetimes is significant \textbf{for MS stars.} Thus determining metallicity
of solar-like stars, especially for the `low-mass' type stars, aids us in understanding
their oscillations.

\begin{figure}
\centering
\includegraphics[scale=0.25]{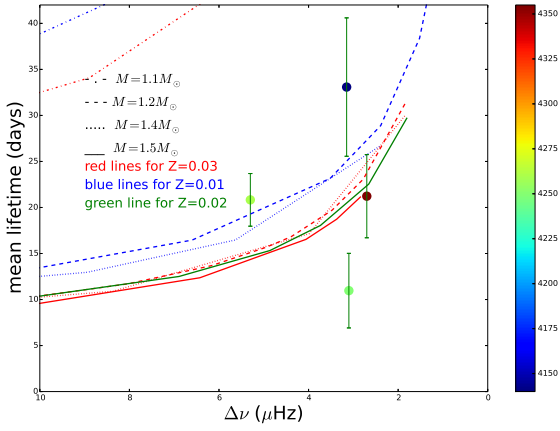}
\caption{ 
Comparison between the mean lifetimes of \cite{hekk10} sample and the
mean lifetimes computed by using Equation (\ref{tau1}) with $\tau_{\odot}=3.8$ days
and $f=3$. The colorbar is proportional to \dteff{} of the sample.
}
\label{fig6}
\end{figure}

Figure \ref{fig6} compares the values of $\tau$ of \cite{hekk10} sample and 
those \textbf{calculated from models by using Equation (\ref{tau1}) with $\tau_{\odot}=3.8$ days.}
The mean lifetimes of \cite{hekk10} sample can be reproduced by \textbf{
models with masses between about $1.2$ and $1.5$ \dsm{}.}

\begin{figure}
\includegraphics[scale=0.2]{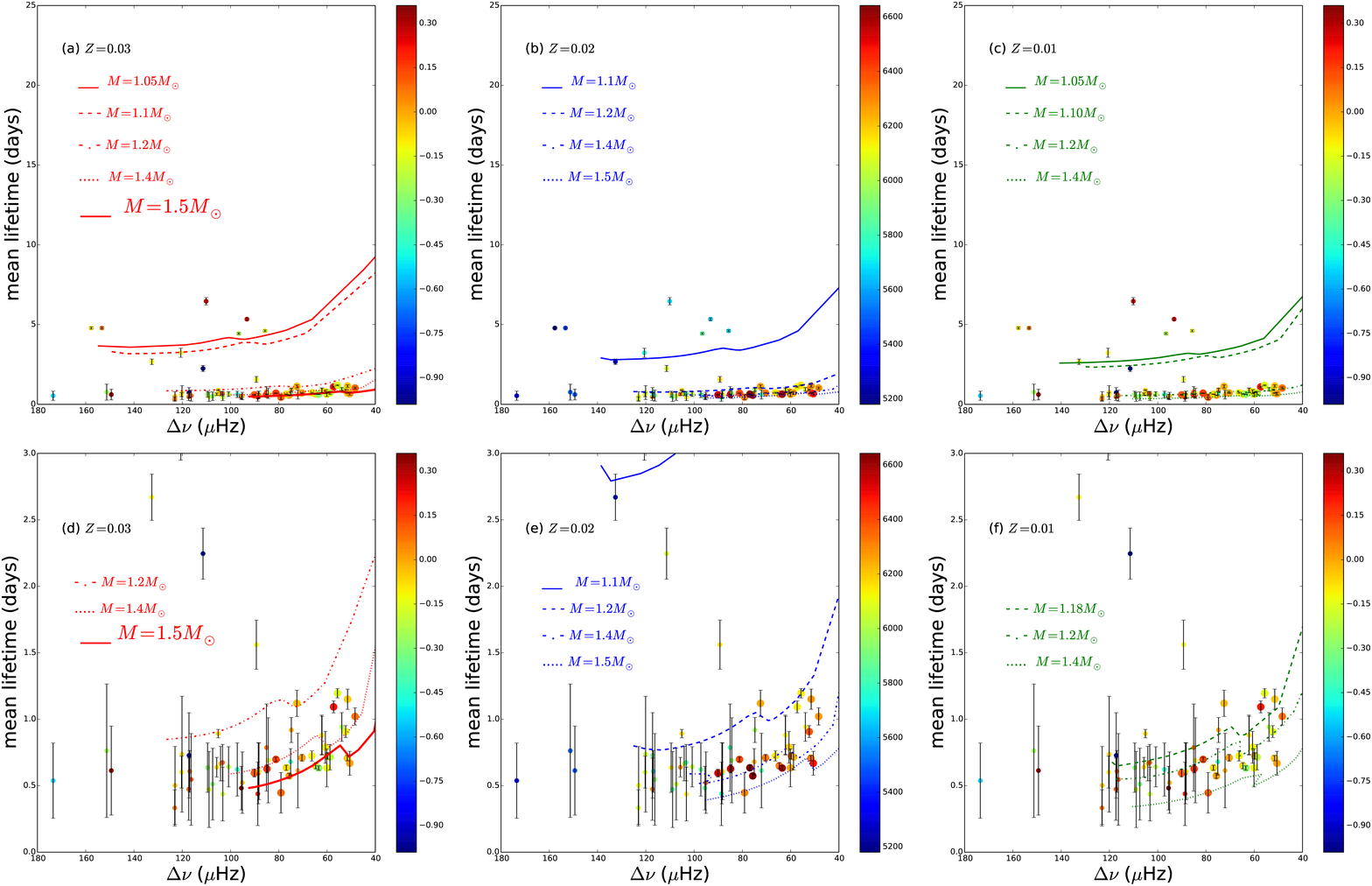}
\caption{ 
Same as Figure \ref{fig4}, but the theoretical mean lifetimes calculated by using 
Equation (\ref{tau2}) with $\tau_{\odot}=3.8$ days. The value of $f_{2}$ is $1$
for models with $M\lesssim1.1$ \dsm{} and $1/3$ for models with $M\gtrsim1.2$ \dsm{}.
}
\label{fig7}
\end{figure}

\textbf{The mean lifetimes of oscillations calculated by using Equation (\ref{tau2}) 
are shown in Figure \ref{fig7}. Compared with the results shown in
Figure \ref{fig4}, Equation (\ref{tau2}) seems to underestimate
the values of $\tau$ of models with $Z=0.03$ and $0.02$. This indicates
that the acoustic impedance is dependent on metallicity. Moreover,
in the calculations for Figure \ref{fig4},
we adoted $f=3$, which may overestimate the values of $\tau$ for more massive 
stars. For these stars, the value of $f$ might be larger than $3$ and should 
be determined from detailed asteroseismic analyses.}

\begin{figure}
\centering
\includegraphics[scale=0.25]{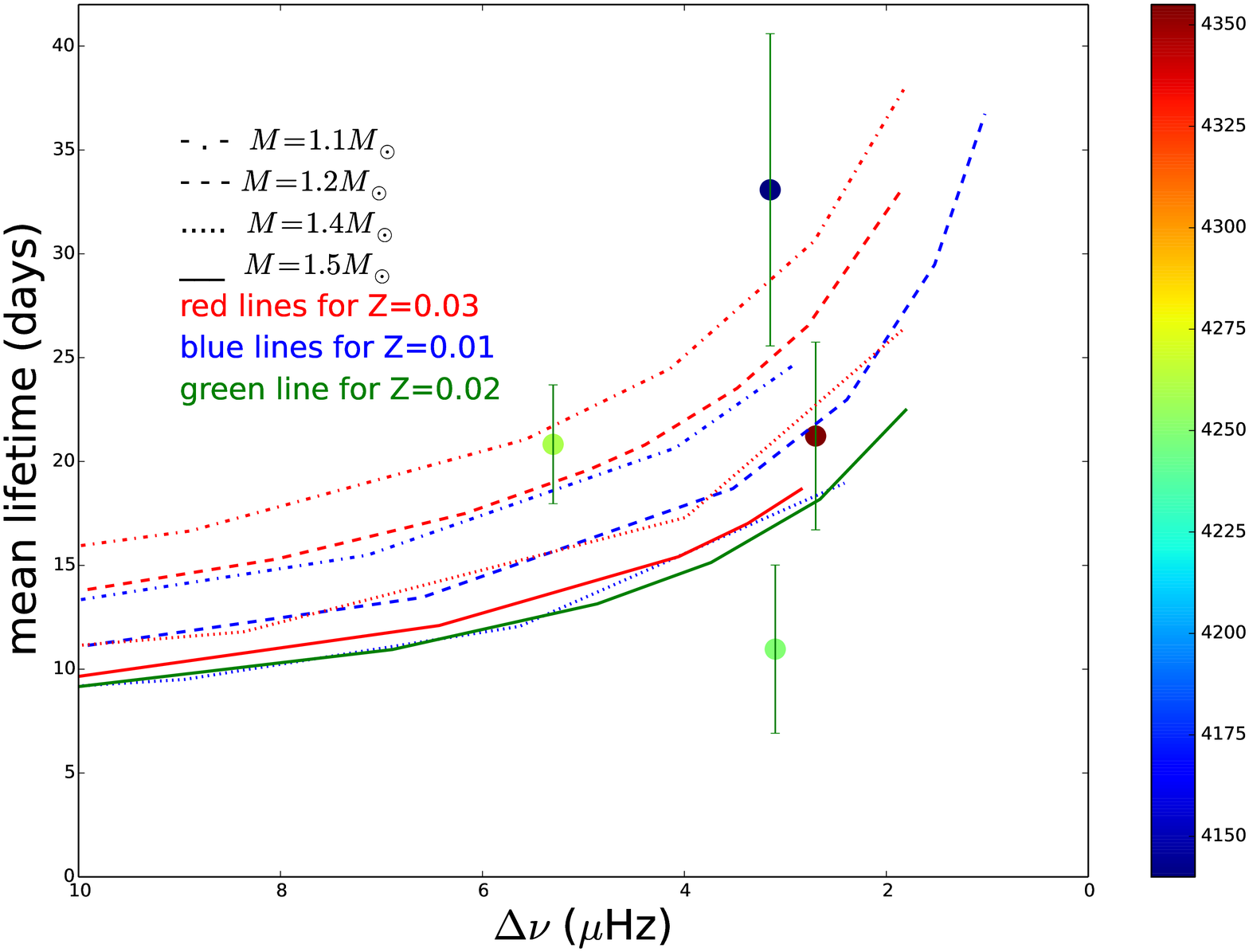}
\caption{ 
Same as Figure \ref{fig7}, but for red giants. The theoretical mean lifetimes were calculated
by using Equation (\ref{tau2}) with $f_{2}=1$.
}
\label{fig8}
\end{figure}

\textbf{For red giants with a given mass and radius, the higher the metallicity,
the lower the effective temperature and $\rho c$. Therefore, compared with 
Equation (\ref{tau1}), Equation (\ref{tau2}) overestimates $\tau$ of stars 
with a higher metallicity (see Figures \ref{fig6} and \ref{fig8}).}

\begin{figure}
\includegraphics[scale=0.2]{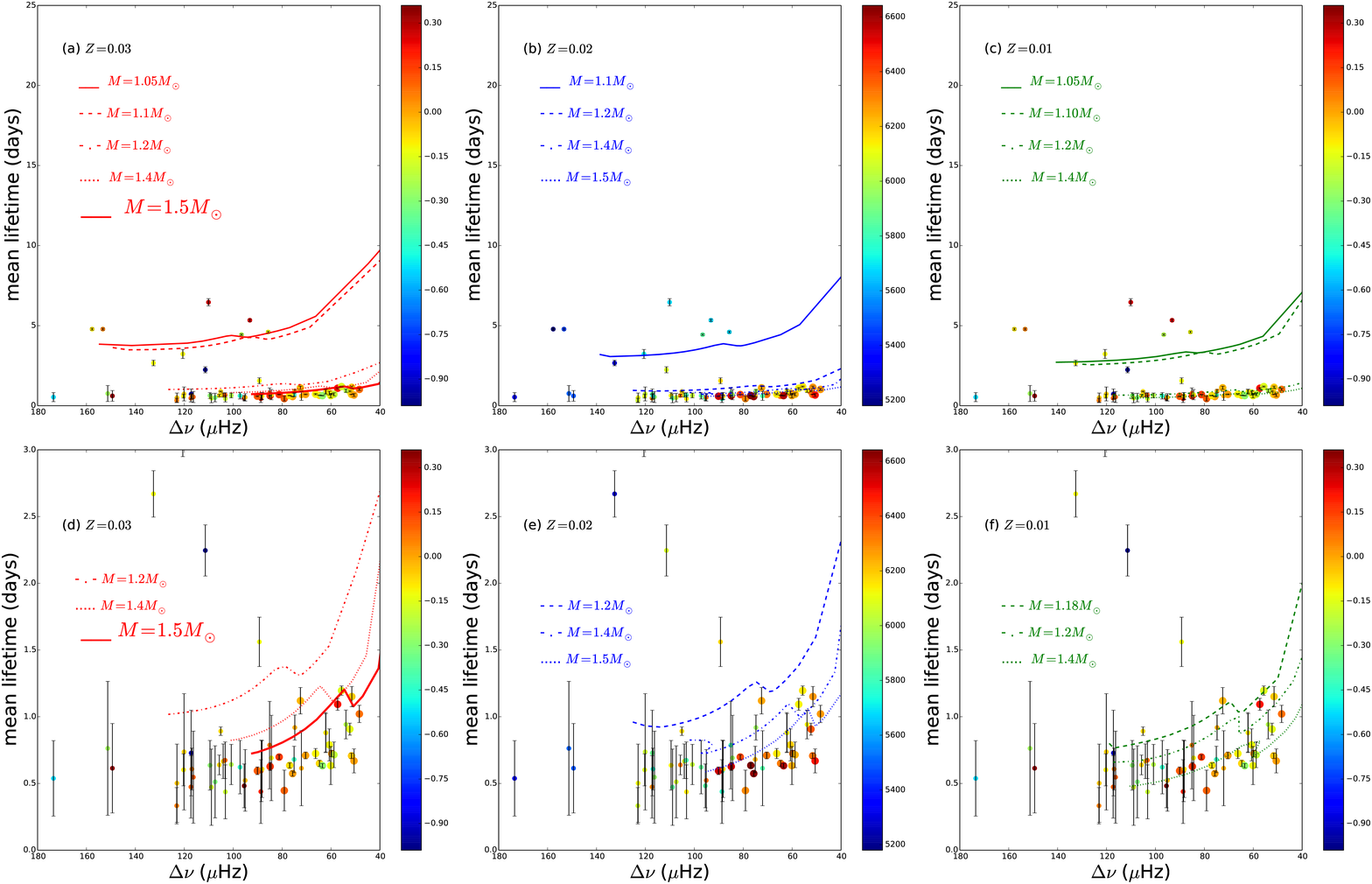}
\caption{ 
Same as Figure \ref{fig4}, but the theoretical mean lifetimes calculated by using 
Equation (\ref{tau3}) with $\tau_{\odot}=3.8$ days. The value of $f_{3}$ is $1$
for models with $M\lesssim1.1$ \dsm{} and $1/3$ for models with $M\gtrsim1.2$ \dsm{}.
}
\label{fig9}
\end{figure}

\begin{figure}
\centering
\includegraphics[scale=0.25]{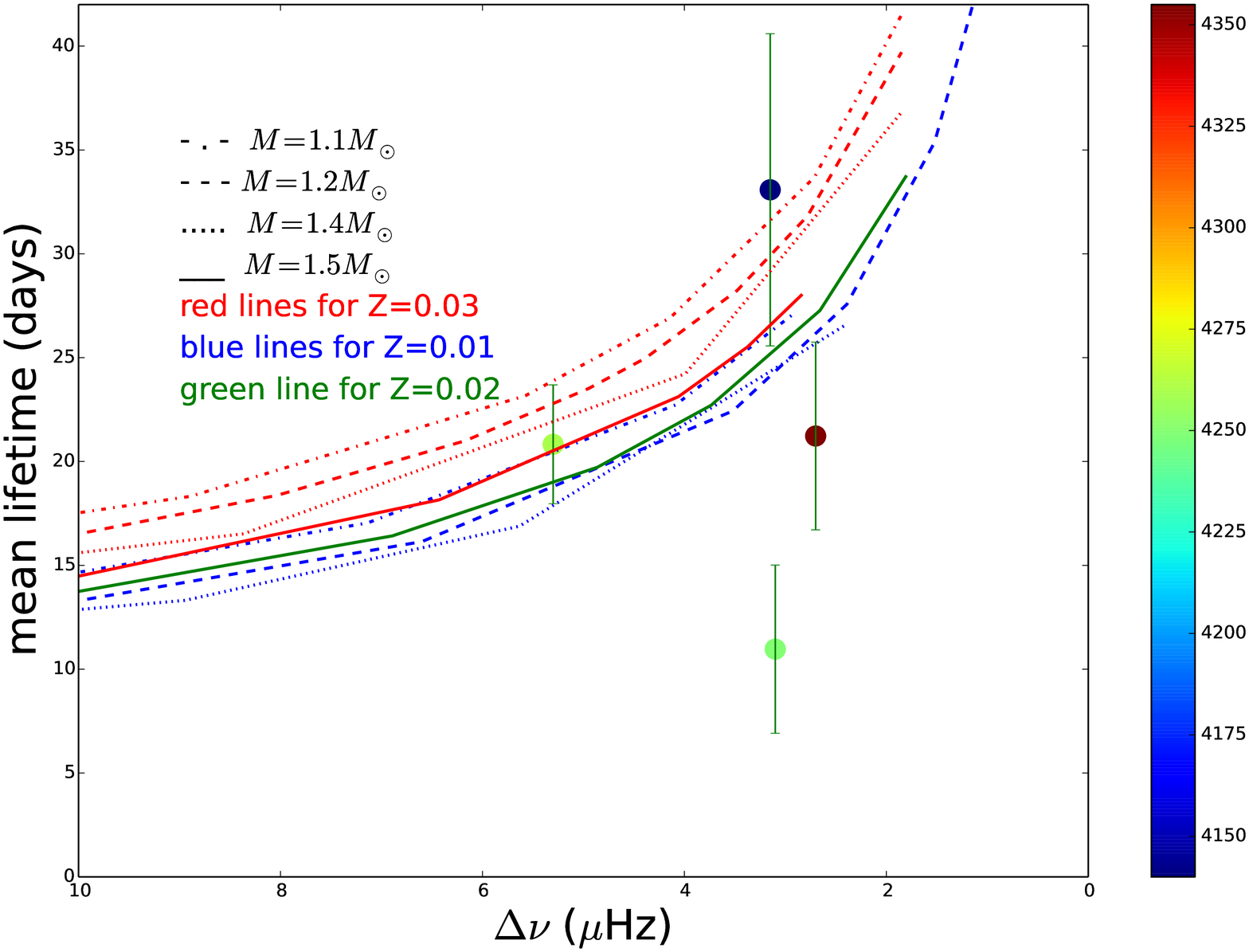}
\caption{ 
Same as Figure \ref{fig9}, but for red giants. The theoretical mean lifetimes were calculated
by using Equation (\ref{tau3}) with $f_{3}=1$.
}
\label{fig10}
\end{figure}

\textbf{Figures \ref{fig9} and \ref{fig10} represent the results calculated by using
Equation (\ref{tau3}), which shows that Equation (\ref{tau3}) is a good approximation
of Equation (\ref{tau2}). However, Equation (\ref{tau3}) slightly overestimates 
the values of $\tau$ of stars with $M\gtrsim1.2$ \dsm{}. Moreover, it 
slightly overestimates the $\tau$ of red giants with $Z=0.03$.}

\section{DISCUSSION AND SUMMARY}
\subsection{Discussion}

For some stars whose \ddn{}, \dteff{}, and $\tau$ have been determined, the acoustic impedance
$\rho c$ at the surface of the stars can be determined by Equation (\ref{tau1}).
\textbf{Thus determining the linewidths of modes with $\nu\sim$ \dnmax{} play an important 
role in asteroseismology.}

\textbf{In the calculations of the results shown in Figure \ref{fig4}, we adoted 
the mean value of $f$. This could overestimate the values of $\tau$ of more massive 
stars and lead to a difference between the theoretical mean lifetimes and the observed ones. 
The dependence of $f$ and $f_{2}$ on mass could be obtained from detailed
asteroseismic analyses of some F-type stars with solar-like oscillations.}

The metallicity of a star can affect the energy transfer in the star. Thus the work integral
of Equation (\ref{we1}) can be affected by metallicity, i.e. $\Gamma$ or $\tau$ of 
modes can be affected by metallicity. The mean lifetime of CoRoT 102767771 in \cite{hekk10}
sample \textbf{is overestimated} by Equation (\ref{tau3}) (see Figure \ref{fig3}). 
This could be related to the effect of metallicity. \textbf{Compared with the results 
of Equation (\ref{tau1}), calculations show that Equation (\ref{tau3}) actually 
overestimates the values of $\tau$ of red giants with $Z_{i}=0.03$ (see Figure \ref{fig10}).
The change in metallicity for a star with a given mass can lead to a variation in 
the effective temperature and radius of the star or a variation in $\rho c$. 
This effect has been considered by the scaling relations. Therefore, the effect 
of metallicity cannot significantly change the results of Equation (\ref{tau1}). }

\textbf{The mean lifetime of oscillations of a star is dependent on the acoustic 
impedance. However, the acoustic impedance is hard to be estimated in observation.}
The value of $\rho c/(\rho c)_{\odot}$ is $1/2$ for `more massive' sample 
and $1/5$ for `red' sample. This indicates that the value of the acoustic
impedance of stars decreases with mass and \textbf{radius. Thus we can 
assume $\rho c \propto 1/(MR)$. The calculations show that the effect of $\rho c$ 
on $\tau$ can be approximated to the effect of $1/(MR)$. Parameters $f_{2}$ 
and $f_{3}$ derive from the assumption of $\rho c \propto 1/(MR)$ and 
the dependence of $f$ on the mass of stars. The calculations show that the values 
of $f_{2}$ and $f_{3}$ are about $1$ for MS stars with $M\lesssim1.1$ \dsm{} 
but are about $1/3$ for MS stars with $M\gtrsim1.2$ \dsm{}. This can be partly 
due to the fact that the value of $(\Gamma_{3}-1)_{R}$ of MS stars with $M\gtrsim1.2$ \dsm{}
is about $3$ times greater than that of the Sun. The values of $f_{2}$ and $f_{3}$ 
are $1$ for red giants. In red-giant phase, the radius of a star with $M=1.2$ \dsm{} can 
increase by more than $10$ times. The rapid change in radius of red giants is the 
main factor affecting the $\tau$ of red giants.
}

\textbf{Figures \ref{fig4} and \ref{fig5} show that there are three stars whose $\tau$ 
is hard to be explained by stellar models. The values of their \ddn{} are larger
than $\sim150$ \dhz{}, which indicates that their masses might be less than $1$ \dsm{},
i.e. the values of their $\rho c$ could be larger than that of the Sun.
But their $\tau$ can be well reproduced by Equation (\ref{tau1}) 
with $\rho c/[3(\rho c)_{\odot}]=1/6$ (see panel $c$ of Figure \ref{fig2}).}

\subsection{Summary}
 
In this work, basing on the works of \cite{balm90} and \cite{gold91} and the 
definition of \cite{unno89}, we deduced a \textbf{formula to describe the linewidth
of the mode with $\nu\sim\nu_{max}$. Due to the fact that the frequencies of 
$p$-modes have an approximately equal separation, the mean linewidth is approximate 
to $\Gamma_{\omega_{max}}$. By using $\tau=1/(\pi\Gamma$), we obtained a} scaling 
relation to describe the average lifetime of solar-like oscillations of stars. 
The \textbf{mean linewidth and lifetime are} determined by the large frequency 
separation, the effective temperature, and the acoustic impedance $\rho c$. 
Furthermore, the dependence of the mean lifetime on the acoustic impedance 
can be \textbf{roughly reduced to a dependence on the mass and radius} of stars. 
However, this will introduce a free parameter 
into the formula of the mean lifetime. \textbf{The calculations show that 
the mean lifetimes of $p$-modes of stars decrease with an increase in mass 
of stars. The mean lifetimes also decrease with a decrease in metallicity
for MS stars. }

We compared the results of the scaling relations with the observational
results of \textit{Kepler} \citep{lund17} and \textit{CoRoT} \citep{hekk10}. Most of
the observational results are well reproduced. This indicates that the scaling relations
work well. Our calculations show that the lifetime of modes can be affected by
stellar metallicity. The effects of metallicity cannot be fully described by the change
in radius and effective temperature. Therefore, the effects of metallicity on $\tau$
could lead to the fact that $\tau$ of some stars deviates from the results of the 
scaling relations, such as CoRoT 102767771.

\section*{Acknowledgments}

\clearpage

\clearpage

\clearpage
\clearpage

\clearpage

\clearpage

\end{document}